\documentclass[12pt]{article}

\newcommand{\bee}{\begin{equation}}
\newcommand{\ee}{\end{equation}}
\newcommand{\bea}{\begin{eqnarray}}
\newcommand{\eea}{\end{eqnarray}}
\newcommand{\R}{\rm I\kern-.2emR}
\newcommand{\C}{\rm \kern.25em\vrule height1.4ex
depth-.12ex width.06em\kern-.31em C}
\newcommand{\N}{{\rm I\kern-.16em N}}
\newcommand{\Z}{{\rm Z\kern-.35em Z}}

\begin{document}                                                                
\begin{flushright}
AZPH-TH-97/01 \\
MPI-PhT 97-022 \\
\end{flushright}
\bigskip\bigskip\begin{center}
{\Huge
Does Conformal Quantum Field Theory Describe the Continuum
Limits of $2D$ Spin Models with Continuous Symmetry?
}                                                                               
\end{center}
\vskip 1.0truecm
\centerline{Adrian Patrascioiu}
\centerline{\it Physics Department, University of Arizona,}
\centerline{\it Tucson, AZ 85721, U.S.A.}
\vskip5mm
\centerline{and}
\centerline{Erhard Seiler}
\centerline{\it Max-Planck-Institut f\"ur Physik}
\centerline{\it (Werner-Heisenberg-Institut)}
\centerline{\it F\"ohringer Ring 6, 80805 Munich, Germany}
\bigskip \nopagebreak \begin{abstract}

It is generally taken for granted that two-dimensional critical 
phenomena can be fully classified by the well known two-dimensional 
(rational) conformal quantum field theories (CQFTs). In particular it
is believed that in models with a continuous symmetry characterized by
a Lie group $G$ the continuum theory enjoys an enhanced symmetry 
$G\times G$ due to the decoupling of right and left movers. In this
letter we review the conventional arguments leading to this conclusion,
point out two gaps and provide a conterexample. Nevertheless we justify
in the end the conventional conclusions by additional arguments.
\end{abstract}
\vskip2mm
Since the works of Belavin, Polyakov and Zamolodchikov \cite{BPZ}
as well as Friedan, Qiu and Shenker \cite{FQS}
it is generally accepted that two-dimensional critical phenomena can
be fully classified by the well-known (rational) conformal Quantum
Field Theories. This is also supposed to provide the explanation
for the rational critical exponents occurring in these models.

In theories with a continuous symmetry group $G$ the conformal philosophy
leads to a `doubling' of the symmetry to $G\times G$ \cite{affleck}, 
with left and right chiral theories both separately invariant under $G$.
Applying this philosophy to the model with the simplest continuous
symmetry, namely the critical $O(2)$ model, it is concluded that
it is completely describable by the Kac-Moody and Virasoro algebras 
associated to the massless free scalar field (see for instance 
\cite{affleck}). This means in particular that the Noether current is a 
gradient, i.e.its {\it curl} vanishes.

In this letter we point out that the conventional arguments contain two 
gaps: firstly, it is not a priori guaranteed that the Noether current
in the continuum exists as a local quantum field, and secondly, it
could turn out that the Noether current is ultralocal in the Euclidean
world, leading to its vanishing in the Minkowski world. For a more
detailed discussion of the issues involved, see \cite{curro2}.

Conventionally, the arguments leading to the splitting of the theory
into two independent `chiral' theories and the ensuing enhancement of 
the symmetry are given in the framework of Minkowski space Quantum
Field Theory. For our purpose here, let us first transcribe those
arguments to the Euclidean setting: assuming a scale invariant continuum 
theory with a conserved current $j_\mu(x)$, Euclidean covariance requires
that the two-point function $G_{\mu\nu}$ of $j_\mu$ is of the form
\bee
G_{\mu\nu}\equiv\langle j_\mu(0) j_\nu(x)\rangle=\delta_{\mu\nu}
{b\over x^2}+{ax_\mu x_\nu\over(x^2)^2} \quad \ (x\neq 0)
\ee
and current conservation requires
\bee
a=-2b.
\ee
This means that $G_{\mu\nu}$ is, up to a factor $b$, equal to the
two point function of $\partial_\mu\phi$ where $\phi$ is the massless 
free scalar field (it is irrelevant here that the massless scalar field 
does not exist as a Wightman field). The two-point function of the dual 
current $\epsilon_{\mu\nu}j_\nu$ is then
\bee
\tilde G_{\mu\nu}\equiv \epsilon_{\mu\lambda}\epsilon_{\rho\nu}
G_{\lambda\rho}=G_{\mu\nu}
\ee
so on the level of the two point function the dual current is also 
conserved. By general properties of local quantum field theory 
(Reeh-Schlieder theorem, see \cite{SW}) it follows that these 
conservation laws hold not only for the two point functions, but for the 
quantum fields $j_\mu$ and $\tilde j_\mu$ themselves. 
Conservation of the two currents $j$ and  $\tilde j$ is 
equivalent to conservation of the two chiral currents $j_\pm$.

So the two conservation laws together imply 
that
\bee
j_\mu=\sqrt{b} \partial_\mu\phi  ,
\ee
where $\phi$ is the massless scalar free field, and also that
\bee
j_\mu=\sqrt{b} \epsilon_{\mu\nu}\partial_\nu\psi  ,
\ee
where $\psi$ is another `copy' of the massless scalar free field.

This argument is certainly correct, but it depends on the {\it existence}
of the Noether currents as Wightman fields, and this is nontrivial and
can actually fail, as the following example shows. This example is the 
two-component free field in $2D$ in the massless limit, which also has a 
global $O(2)$ invariance. It is well known that the massless limit makes 
sense only for functions of the gradients of the fields. But the Noether 
current of the $O(2)$ symmetry cannot be written as a function of the 
gradients. It is also easy to see directly that its correlation functions
do not have a limit as $m\to 0$ (see \cite{curro2}).

The two-point function of the {\it curl} of the current of the free field
is proportional to $1/(x^2)^2$ in the massless limit, and if one tries 
to `integrate' this to obtain the current two-point function, one
obtains (imposing Euclidean covariance)
\bee
G_{\mu\nu}(x)=-\delta_{\mu\nu}{\ln x^2+\lambda \over 8x^2}
+ x_\mu x_\nu {\ln x^2+1+\lambda\over 4x^2}
\ee
with a free parameter $\lambda$. But this is not the Euclidean two-point 
function of a Wightman field, because it changes sign as we vary $x$ and 
therefore violates reflection positivity \cite{OS}. So here we have a
case where both the {\it curl} and the divergence of the current exist
as local fields (the divergence vanishes in Minkowski space), but the 
current itself does not. This situation is an instance of `nontrivial 
local cohomology' \cite{strocchi,pohlmeyer,roberts}. So the first gap in 
the conventional argument is the assumption that the local cohomology is 
trivial.

We proceed now to show that in a continuum model with a critical point 
at $\beta_{crt}<\infty$ with a continuous symmetry the situation of the 
counterexample (`nontrivial local cohomology') cannot arise. For the sake
of definiteness we will concentrate on the $O(2)$ model, but it is 
straightforward to generalize the argument.

The $O(2)$ model is determined by its standard Hamiltonian (action)

\bee
H=-\sum_{\langle ij\rangle} s(i)\cdot s(j)
\ee
where the sum is over nearest neighbor pairs on a square lattice
and the spins $s(.)$ are unit vectors in the plane $\R^2$.
As usual Gibbs states are defined by using the Boltzmann factor
$\exp(-\beta H)$ together with the standard a priori measure on the
spins first in a finite volume, and then taking the thermodynamic
limit.

The most interesting property of the model is its so-called KT transition
\cite{KT}, from a high temperature phase with exponential clustering
to a low temperature one with only algebraic decay of correlations.
A recent estimate for the transition point is \cite{Marcu}

\bee
\beta_{KT}=1.1197
\ee

This number is not exact, but here all that matters is
that the correlation length is so large that on the lattices we
simulate it may be treated as infinite.
The nature of the transition is supposed to be peculiar, with exponential
instead of the usual power-like singularities, but this is not our
concern here. Instead we want to study the model at its transition
point. We are in particular interested in the correlations of the
Noether current, given by
\bee
j_\mu(i)=\beta\sin\Bigl(\phi(i+\hat\mu)-\phi(i)\Bigr)
\ee
where
\bee
s_1(i)=\cos\bigl(\phi(i)\bigr), s_2(i)=\sin\bigl(\phi(i)\bigr)
\ee

On a torus the current can be decomposed into 3 pieces, a longitudinal 
one, a transverse one and a constant (harmonic) piece. This 
decomposition is  easiest in momentum space, and effected by the 
projections 
\bee
P^T_{\mu\nu}=\Biggl(\delta_{\mu\nu}-{(e^{ip_\mu}-1)(e^{-ip_\nu}-1)\over
\sum_\alpha(2-2\cos p_\alpha)}\Biggr)(1-\delta_{p0}) ,
\ee

\bee
P^L_{\mu\nu=}={(e^{ip_\mu}-1)(e^{-ip_\nu}-1)\over
\sum_\alpha(2-2\cos p_\alpha)}(1-\delta_{p0})
\ee
and
\bee
P^h_{\mu\nu}=\delta_{\mu\nu}\delta_{p0}.
\ee
with $p_\mu=2\pi n_\mu/L$, $n_\mu=0,1,2,...,L-1$.

We are mainly interested in the tranverse momentum space 2-point function
\bee
F^T(p_1,0)=\langle |\hat j_2(p_1,0)|^2\rangle
\ee
(for $p_1\neq 0$; the hat denotes the Fourier transform);
the longitudinal two-point function
\bee
F^L(p_1,0)=\langle |\hat j_1(p_1,0)|^2\rangle
\ee
is a constant because of current conservation. The constant is determined
by a Ward identity to be $\beta E$ where 
$E=\langle s(0)\cdot s(\hat\mu)\rangle$ (see \cite{curro2}).

The thermodynamic limit is obtained by sending $L\to\infty$ for fixed
$p=2\pi n/L$, so that in the limit $p$ becomes a continuous variable
ranging over the interval $[-\pi,\pi)$. The $O(2)$ model not only does
not show spontaneous symmetry breaking according to the Mermin-Wagner
theorem, but it has a unique infinite volume limit, as shown long ago by
Bricmont, Fontaine and Landau \cite{BFL}. The convergence to the 
thermodynamic limit is seen clearly in the Monte-Carlo simulations
reported in \cite{curro2}.

The continuum limit in the infinite volume is obtained as follows:
let $\hat F(p;\infty)\equiv \hat T(p)$ be the Fourier transform of
the one-dimensional lattice function $T(n)$. In general $\hat T$
has to be considered as a distribution on $[-\pi,\pi)$, and it can be 
extended to a periodic distribution on the whole real line. The 
continuuum limit of $T(n)$ also has to be considered in the sense of 
distributions; it is obtained by introducing an integer $N$ as the unit 
of length, making the lattice spacing equal to $1/N$. For an arbitrary 
test function $f$ (infinitely differentiable and of compact support) on 
the real axis we then have to consider the limit
$N\to\infty$  of
\bee
(T,f)_N\equiv \sum_n f({n\over N}) T(n).
\ee
It is not hard to see that the right hand side of this is equal to 
\bee
{1\over 2\pi}\int_\infty^\infty dq \hat T({q\over N})\hat f(q).
\ee
and converges as $N\to\infty$ to
\bee
{1\over 2\pi} \hat T(0) \int dq \hat f(q)
={1\over 2\pi}f(0)\hat T(0),
\ee
provided $lim_{p\to 0}\hat T(p)\equiv \hat T(0)$ exists, expressing the
fact that in this case the limit of $T$ is a pure contact term.

We want to prove rigorously that the continuum limit of the
thermodynamic limits $\hat F^T(p,\infty)$ and $\hat F^L(p,\infty)$ of
$\hat F^T(p,L)$ and $\hat F^L(p,L)$ are constants; the second fact is 
of course again just a restatement of the Ward identity (12), whereas the
first one expresses the vanishing of {\it curl j} in the continuum, thus 
confirming Affleck's claim regarding the enhancement of the continuous 
symmetry.

We use reflection positivity (RP) of the Gibbs measure formed with
the standard action (see for instance \cite{OSe}) on the periodic
lattice. RP applied to the current two-point functions yields:
\bee
F^L(x_1,L)=\sum_{x_2} \langle j_1(x_1,x_2) j_1(0,0)\rangle\leq 0
\ee 
for $x_1\neq 0$ and
\bee
F^T(x_1,L)=\sum_{x_2} \langle j_2(x_1,x_2) j_2(0,0)\rangle\geq 0  
\ee
for all $x_1$.
From these two equations it follows directly that
\bee
0\leq \hat F^T(p,L)\leq \hat F^T(0,L)=\hat F^L(0,L)\leq \hat F^L(p,L)
=\beta E
\ee 

These inequalities remain of course true in the thermodynamic limit,
but we have to be careful with the order of the limits. If we define
$\hat F^T(p,\infty)$ and $\hat F^L(p,\infty)$ as the Fourier transforms
of $\lim_{L\to\infty} F^T(x,L)$ and $\lim_{L\to\infty} F^L(x,L)$,
respectively, one conclusion can be drawn immediately:

{\it Proposition:}
$\hat F^T(p,\infty)$ and $\hat F^L(p,\infty)$ are continuous functions
of $p\in [-\pi,\pi)$.

The proof is straightforward, because due to the inequalities (19), (20)
and (21) together with the finiteness of $\beta_{KT}$ the limiting
functions $F^L$ and $F^T$ in $x$-space are absolutely summable.

But it is not assured that the limits $L\to\infty$ and $p\to 0$ can
be interchanged, nor that the thermodynamic limit and Fourier
transformation can be interchanged. On the contrary, the numerics
presented in \cite{curro2}, as well as finite size scaling arguments
(see below) suggest that
\bee
\lim_{p\to 0}\lim_{L\to\infty} \hat F^L(p,L)>
\lim_{L\to\infty}\hat F^L(0,L)
\ee
and therefore also
\bee
\lim_{p\to 0}\lim_{L\to\infty} \hat F^L(p,L)
>\lim_{p\to 0}\lim_{L\to\infty} \hat F^T(p,L).
\ee

This fact plays an important role in closing the second gap in Affleck's
arguments. But first we want to show the following:

{\it Proposition:}
In the continuum limit both $\hat F^L(p,\infty)$ and $\hat F^T(p,\infty)$
($p\neq 0$) converge to constants.

{\it Proof:} The proof is essentially contained in eqs.(17) and (18).
We only have to notice that due to eq.(21) 
$\lim_{p\to 0} \hat F^L(p,\infty)$ and $\lim_{p\to 0} \hat F^T(p,\infty)$
exist.

In spite of this result, Affleck's claim could still fail in a different
way if $\hat F^T(p,\infty)$ and $\hat F^L(p,\infty)$ converged to the
same constant in the continuum limit. Let us denote the continuum limit  
of $\hat F^T(p,\infty)$ by $g$. Then the current-current correlation
in this limit is
\bee
\langle j_\mu j_\nu\rangle\hat(p) = \beta E P^L_{\mu\nu}+
g P^T_{\mu\nu}=g\delta_{\mu\nu}+(\beta E-g){p_\mu p_\nu\over p^2}
\ee
So we see that if $g=\beta E$, the current-current correlation reduces to
a pure contact term and vanishes in Minkowski space. Above we proved 
only that
\bee
g\leq \beta E
\ee
This gap in the conventional arguments will be closed using numerical 
simulation data together with finite size scaling arguments; we will show
thereby that the continuum limit $g$ of $\hat F^T(p,\infty)$ is not
equal to $\beta E$.
For $\beta<\beta_{KT}$ the current two point function is decaying
exponentially, hence its Fourier transform is continuous (and even real
analytic). The same applies then to the longitudinal and transverse parts
$\hat F^L(p,\infty)$ and $\hat F^T(p,\infty)$; in particular
\bee
\hat F^T(0,\infty)=\hat F^L(0,\infty)=\beta E
\ee
by the Ward identity mentioned above.

That does not, however, imply that at $\beta=\beta_{KT}$
$\lim_{p\to 0} \hat F^T(p,\infty)=\beta E$, because the current two-point
function cannot be expected to be absolutely summable there. On the
contrary, if we can find that
\bee
\lim_{L\to\infty}\hat F^L(0,L)<\beta E
\ee
this implies also
\bee
g=\lim_{p\to 0}\hat F^T(p,\infty)<\beta E
\ee
because by eq.(22) $\hat F^L(p,\infty)\leq\lim_{L\to\infty}\hat F^L(0,L)$.

The MC data taken at $\beta_{KT}$ and listed in Tab.1 indicate that
\bee
d\equiv\beta E-\hat F^L(0,L)=\hat F^L({2\pi\over L},L)-\hat F^L(0,L)
\ee
goes to a positive number ($<.68$ but probably $>.6$),
suggesting that indeed $g<\beta E$. But the question is whether this
`discontinuity' is a finite volume artefact or not. To address this issue
we took data at $\beta<\beta_{KT}$ keeping the ratio $L/\xi$ fixed while  
increasing $\xi$. In this approach the massless continuum limit would
correspond to $L/\xi\to 0$ (while $L/\xi\to\infty$ would correspond to
the massive continuum limit in a thermodynamic box). Actually we use
$L/\xi_{eff}$ instead of $L/\xi$ as an independent variable, where  
$\xi_{eff}$ is the effective correlation length measured on the lattice
of size $L$; in the finite size scaling limit this is equivalent,
because $L/\xi_{eff}$ becomes a unique monotonic function of $L/\xi$.
The data listed in Tab.2 indicate that $d(L)=\beta E-\hat F^L(p,L)$
(the `discontinuity' of $\hat F^L(p,L)$ at $p=0$) depends only on
$L/\xi_{eff}$ in agreement with finite size scaling, and that it goes to
a number above .6 in the massless continuum limit which is reached around
$L/\xi_{eff}\approx 1.3$. The two sets of Monte-Carlo data together 
say that $\lim_{L\to\infty} d(L)$ is a number between .6 and .68, and
might actually be equal to $2/\pi$. In any case, they provide convincing 
evidence that the Noether current is not an ultralocal field.

A.P is grateful to the Alexander von Humboldt Foundation for a Senior
U.S.Scientist Award and to the Max-Planck-Institut for its hospitality;
E.S. is grateful to the University of Arizona for its hospitality
and financial support.



\noindent
{\bf Tab.1:} {\it The `discontinuity' $d(L)=\beta E-g(L)$
at $\beta_{KT}$ for different values of $L$.}

\begin{tabular}[t]{r|r|r|r}
$ L $ & $L/\xi_{eff}$ & $g(L)$ & $d(L)$   \\
\hline
\hline
25    & 1.2575  & .09769       & .7117(10) \\
50    & 1.2657  & .10666       & .7027(18)  \\
100   & 1.2667  & .11638       & .6930(11)  \\
200   & 1.2722  & .12357       & .6858(08)   \\
400   & 1.2839  & .12944       & .6799(16)
\end{tabular}
\vskip4mm
\noindent
{\bf Tab.2:} {\it The `discontinuity' $d(L)=\beta E-g(L)$
at various values of $\beta<\beta_{KT}$ and $L$.}

\begin{tabular}[t]{r|r|r|r}
$ \beta $ & $ L $ & $ L/\xi_{eff} $ & $d(L)$   \\
\hline
\hline
.93 & 12  & 1.8319  &  .3646(22) \\
.93 & 24  & 2.4441  &  .1926(24) \\
.93 & 36  & 3.2259  &  .0875(34) \\
\hline
.96 & 18  & 1.8314  &  .3636(24) \\
.96 & 36  & 2.4242  &  .1929(31) \\
.96 & 54  & 3.1996  &  .0926(34) \\
\hline
.99 & 32  & 1.8595  &  .3543(48) \\
.99 & 63  & 2.4537  &  .1919(64) \\
.99 & 64  & 2.4467  &  .1862(46) \\
.99 & 96  & 3.2110  &  .0909(43) \\
\hline
1.04 & 63 & 1.5868  &  .4709(59) \\
\hline
1.06 & 63 & 1.4549  &  .5411(49) \\
\hline
1.08 & 63 & 1.3722  &  .6012(36) \\
1.08 & 126& 1.4188  &  .5818(46) \\
\hline
1.09 & 126& 1.3990  &  .6174(18)

\end{tabular}

\end{document}